\pdfoutput=1

\documentclass[fleqn,usenatbib]{mnras}

\usepackage{newtxtext,newtxmath}
\usepackage[T1]{fontenc}

\DeclareRobustCommand{\VAN}[3]{#2}
\let\VANthebibliography\thebibliography
\def\thebibliography{\DeclareRobustCommand{\VAN}[3]{##3}\VANthebibliography}

\usepackage{graphicx}	% Including figure files
\usepackage{rotating}
\usepackage{amsmath}	% Advanced maths commands
\title[A targeted search for Main Belt Comets]{A targeted search for Main Belt Comets}

\author[]{Léa Ferellec$^{1}$,\thanks{E-mail: lea.ferellec@ed.ac.uk}
Colin Snodgrass$^{1}$, 
Alan Fitzsimmons$^{2}$,
Agata Rożek$^{1}$,
Daniel Gardener$^{1}$ \newauthor 
Richard Smith$^{2}$, 
Hissa Medeiros$^{3,4}$,
Cyrielle Opitom$^{1}$,
Henry H. Hsieh$^{5}$\\
$^{1}$Institute for Astronomy, University of Edinburgh, Royal Observatory, Edinburgh, EH9 3HJ, UK\\
$^{2}$Astrophysics Research Centre, School of Physics and Astronomy, Queen’s University Belfast, Belfast BT7 1NN, UK\\
$^{3}$Instituto de Astrofísica de Canarias (IAC), Calle Vía Láctea s/n, 38205 La Laguna, Spain\\
$^{3}$Departamento de Astrofísica, Universidad de La Laguna, Calle Padre Herrera, s/n, 38200 La Laguna, Spain\\
$^{5}$Planetary Science Institute, 1700 East Fort Lowell Rd., Suite 106, Tucson, AZ 85719, USA
}

\date{Accepted XXX. Received YYY; in original form ZZZ}

\pubyear{2022}

\begin{document}
\label{firstpage}
\pagerange{\pageref{firstpage}--\pageref{lastpage}}
\maketitle

\begin{abstract}

Main Belt Comets (MBCs) exhibit sublimation-driven activity while occupying asteroid-like orbits in the Main Asteroid Belt. MBCs and candidates show stronger clustering of their longitudes of perihelion around 15$^{\circ}$ than other objects from the Outer Main Belt (OMB). This potential property of MBCs could facilitate the discovery of new candidates by observing objects in similar orbits.
We acquired deep r-band images of 534 targeted asteroids using the INT/WFC between 2018 and 2020. Our sample is comprised of OMB objects observed near perihelion, with longitudes of perihelion between 0$^{\circ}$ and 30$^{\circ}$ and orbital parameters similar to knowns MBCs. Our pipeline applied activity detection methods to 319 of these objects to look for tails or comae, and we visually inspected the remaining asteroids. 
Our activity detection pipeline highlighted a faint anti-solar tail-like feature around 2001 NL19 (279870) observed on 2018 November 07, six months after perihelion. This is consistent with  cometary activity but additional observations of this object will be needed during its next perihelion to investigate its potential MBC status. If it is active our survey yields a detection rate of $\sim$1:300, which is higher than previous similar surveys, supporting the idea of dynamical clustering of MBCs. If not, it is consistent with previously estimated abundance rates of MBCs in the OMB ({<1:500}).

\end{abstract}

\begin{keywords}
 minor planets, asteroids: general -- comets: general
\end{keywords}

%%%%%%%%%%%%%%%%%%%%%%%%%%%%%%%%%%%%%%%%%%%%%%%%%%

%%%%%%%%%%%%%%%%% BODY OF PAPER %%%%%%%%%%%%%%%%%%

\section{Introduction}

\subsection{Main Belt Comets} Main Belt Comets (MBCs) display comet-like outgassing while occupying asteroidal orbits in the Main Asteroid Belt. Asteroids and comets have long been perceived as two very distinct populations, initially formed within and beyond the snowline, and existing in different reservoirs. The recent discovery of icy objects in the Main Belt, as well as of other active asteroids and inactive comets, blurs the physical and dynamical boundaries between comets and asteroids and indicates it is rather a continuum \citep{JewittHsieh2022}.

MBCs constitute a subset of the active asteroids along with disrupted asteroids, the activity of which is not due to ice sublimation but rather other phenomena such as impacts or rotational effects.
To categorize an active asteroid as an MBC, possible evidence for outgassing includes recurring activity during multiple passages at perihelion, strong non-gravitational acceleration or prolonged dust emission. No direct detection of gas has been made so far due to the low gas production rates in MBCs \citep{Snodgrass2017}.

The first known MBC 133P/Elst-Pizzaro was discovered in 1996 \citep{Elst}. There are now 9 objects in total that have shown recurring activity near perihelion and around 20 additional candidates are known (Table \ref{tab:knownMBCs}). \citet{Sonnett2011} estimated the average MBC to asteroid ratio to be 1:400 in the Main Belt but only a small number will be active at any given time. Efforts have been deployed to discover MBCs in archival data or through targeted or untargeted surveys, but their scarcity and transient activity, in addition to observational constraints, make it challenging. Attempts at finding active objects in archival data from the the Canada–France–Hawaii Telescope Legacy Survey \citep{GILBERT2009,GILBERT2010}, the Thousand Asteroid Light Curve Survey (TALCS) \citep{Sonnett2011}, the Palomar Transient Factory survey \citep{Waszczak} or the Hyper-Suprime-Cam/Suprime-Cam data \citep{Hsieh2016,Schwamb2017} have not lead to any new detection of MBCs.

\begin{table*}
\centering
\caption{
List of known MBCs and candidates, their orbital characteristics and likely source of activity. \textbf{S}: Sublimation. \textbf{R}: Rotation. \textbf{I}: Impact. (\textbf{*}): Objects considered by \citet{kim}. (\textbf{**}):  {Objects that have shown recurring activity near perihelion.}  Objects with a>2.82 belong to the OMB.}
\label{tab:knownMBCs}
\begin{tabular}{lccccccc}
\hline
\textbf{Object} & \textbf{Origin of activity} & \textbf{a} (AU) & \textbf{e} & \textbf{i} (deg) & \textbf{$T_{\mathrm{J}}$} & \textbf{P} (yr) & \textbf{$\bar{\omega}$} (deg) \\ \hline
1 Ceres & S & 2.77  & 0.078  & 10.6  & 3.310     & 4.6  &  154.0         \\  \hline
133P/Elst–Pizarro (P/1996 N2) (*) {(**)}  & S & 3.16  & 0.157  & 1.4  & 3.184     & 5.64  &  -68.0       \\  \hline
176P/LINEAR (*)  & S/I? & 3.19  & 0.193  & 0.2  & 3.167     & 5.72  &  21.0         \\  \hline
233P/La Sagra (P/2005 JR71) & S & 3.03  & 0.411  & 11.3  & 3.081     & 5.28  &  102.2         \\  \hline
238P/Read (P/2005 U1) (*) {(**)}  & S & 3.16  & 0.253  & 1.3  & 3.154     & 5.61  &  17.6       \\  \hline
259P/Garradd (P/2008 R1)  {(**)} & S & 2.73  & 0.342  & 15.9  & 3.217     & 4.52  &  -51.0        \\  \hline
288P (2006 VW139)(*) {(**)}  & S & 3.05  & 0.201  & 3.2  & 3.203     & 5.34  &  3.2      \\  \hline
313P/Gibbs (P/2003 S10) (*) {(**)}  & S & 3.15  & 0.242  & 11.0  & 3.133     & 5.61  &  0.0      \\  \hline
324P/La Sagra (P/2010 R2) (*) {(**)}  & S & 3.09  & 0.154  & 21.4  & 3.100     & 5.45  &  -30.1        \\  \hline
358P/PANSTARRS (P/2012 T1) (*) {(**)}  & S & 3.16  & 0.236  & 11.1  & 3.134     & 5.61  &  26.8        \\  \hline
427P (P/2017 S5) (ATLAS) & S/R? & 3.17  & 0.313  & 11.8  & 3.092     & 5.64  &  -8.1       \\  \hline
432P (P/2021 N4) & ? & 3.04  & 0.243  & 10.1  & 3.173     & 5.28  &  -16.0        \\  \hline
433P (248370) (2005 QN173)  {(**)} & S/R/I? & 3.07  & 0.226  & 0.1  & 3.192     & 5.37  &  -40.0        \\  \hline
P/2016 J1-A (PANSTARRS) (*) & S/I/R? & 3.17  & 0.228  & 14.3  & 3.113     & 5.64  &  -113.4        \\  \hline
P/2013 R3 (Catalina-PANSTARRS) (*) & S/R? & 3.03  & 0.273  & 0.9  & 3.184     & 5.28  &  -8.77 \\ \hline

P/2015 X6 (PANSTARRS) & R/S? & 2.75  & 0.17  & 4.6  & 3.319     & 4.57  &  76.0         \\  \hline

P/2017 S9 (PANSTARRS) & ? & 3.16  & 0.305  & 14.1  & 3.087     & 5.61  &  24.0        \\  \hline
P/2018 P3 (PANSTARRS)  {(**)} & S & 3.01  & 0.416  & 8.9  & 3.096     & 5.2  &  5.2         \\  \hline
P/2019 A3 (PANSTARRS) & ? & 3.15  & 0.265  & 15.4  & 3.099     & 5.59  &  -3.7        \\  \hline
P/2019 A4 (PANSTARRS) &  {R/I}? & 2.61  & 0.09  & 13.3  & 3.365     & 4.22  &  101.0       \\  \hline
P/2019 A7 (PANSTARRS) & ? & 3.19  & 0.161  & 17.8  & 3.103     & 5.69  &  41.0     \\  \hline
P/2020 O1 (Lemmon-PANSTARRS) &  {S/R?} & 2.65  & 0.12  & 5.2  & 3.376     & 4.3  &  -79.0        \\  \hline
P/2021 A5 (PANSTARRS) &  {S?} & 3.05  & 0.14  & 18.2  & 3.147     & 5.31  &  30.0       \\  \hline

P/2021 L4 & ? & 3.17  & 0.119  & 17.0  & 3.125     & 5.64  &  118.0 \\ \hline
P/2021 R8 & ? & 3.02  & 0.294  & 2.2  & 3.179     & 5.26  &  -2.0 \\ \hline
\end{tabular}

\end{table*}

%______________________________
\subsection{Orbital clustering of known MBCs}
\label{sec:kim}

\begin{figure}
    \centering
    \includegraphics[width=\linewidth]{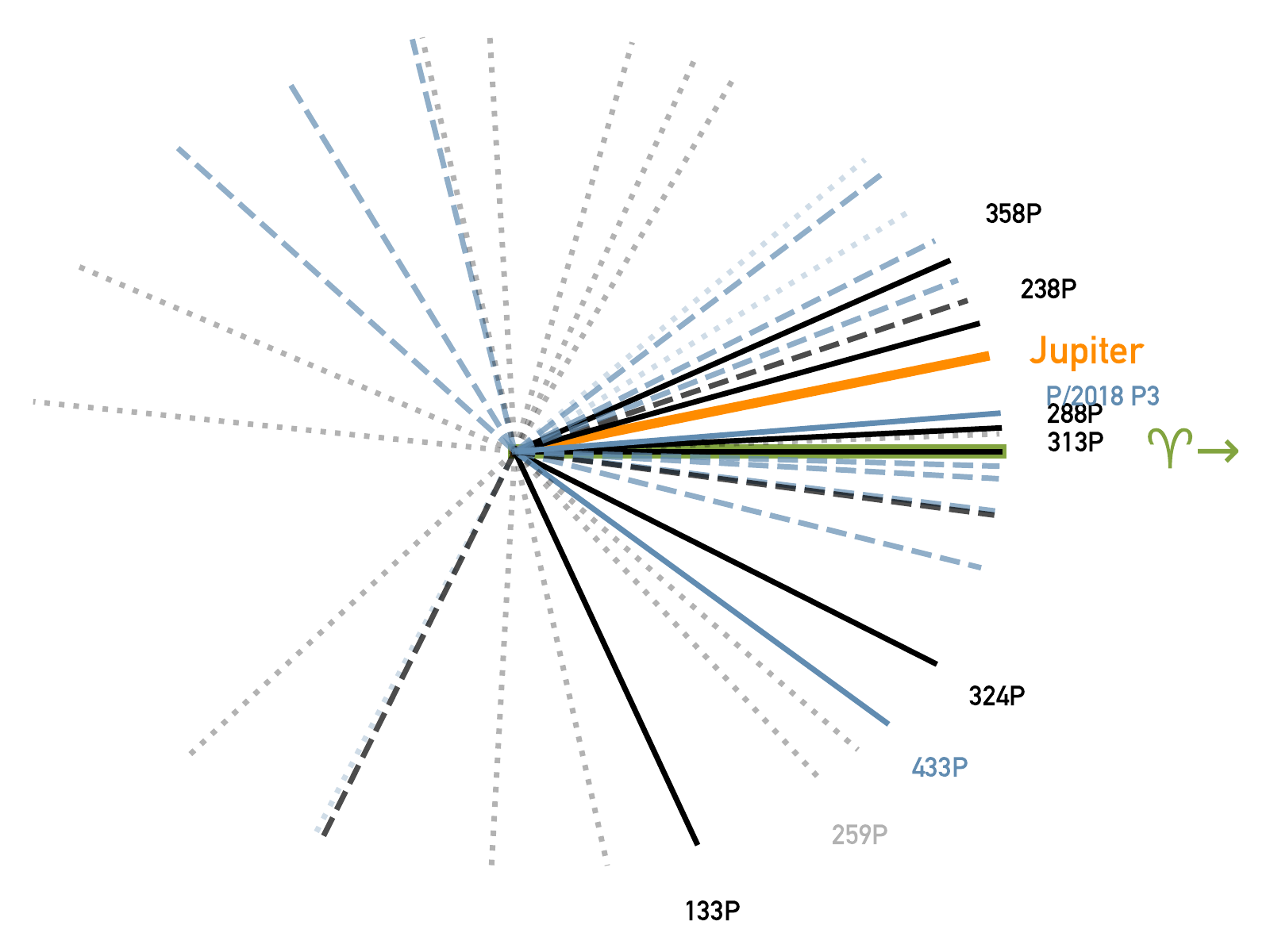}
    \caption{Longitudes of perihelion of known MBCs (solid lines) and MBC candidates (dashed lines). Objects considered by \citet{kim} are shown in black. New additions to the population are shown in blue. Dotted lines represent non-OMB MBCs and other active asteroids. The names of objects that have shown recurring activity near perihelion are shown. The green line indicates the direction of the vernal equinox ($0^{\circ}$).}
    \label{fig:kimetalupdate}
\end{figure}

\citet{kim} analysed orbital elements of confirmed and suspected MBCs of the Outer Main Belt (OMB), and found that the longitudes of perihelion $\bar{\omega}$ of these objects seem to cluster around that of Jupiter ($\bar{\omega}_{\mathrm{J}} \approx 15^{\circ}$), as illustrated by Fig. \ref{fig:kimetalupdate}. Rayleigh's Z-test measures the probability that a sample of circular data (e.g. angles) is drawn from a uniform distribution. Using Rayleigh's Z-test the authors found that the probability that these longitudes were drawn from a uniform distribution was $<0.3$ per cent. Such a clustering is a known feature of the whole OMB population, which is more affected by Jupiter dynamically, but they used Watson's U$^2$ test to show that the clustering was stronger among MBCs and MBC candidates. Watson's U$^2$ test measures the probability that two samples of circular data are drawn from the same distribution. They found that the probability for the longitudes of perihelion of the MBC sample to be drawn from the same distribution as the longitudes of perihelion of all OMB objects was $<2$ per cent.

\begin{table}
\centering
\caption{Results of statistical tests from \citet{kim} and this study. Rayleigh's Z-test measures the probability that the sample ({here the longitudes of perihelion of MBCs and active asteroids from the OMB for which ice sublimation is a possible driver of activity}) is drawn from a uniform distribution. Watson's U$^2$ test measures the probability that two samples (here the previous sample and the longitudes of perihelion of all known OMB objects) are drawn from the same distribution.}
\begin{tabular}{lcc}
\hline
            & \citet{kim} & This study (2022) \\ \hline 
Rayleigh's Z-test   &  $<0.3\%$    &   $<0.001\%$          \\ \hline
Watson's U$^2$ &   $<2\%$  &   $<1\%$         \\ \hline
\end{tabular}
\label{tab:statkim}
\end{table}

\citet{kim} accounted for 9 asteroids, among which only 5 had shown recurring activity. Since then, 3 new MBCs were identified in the Outer Main Belt along with many candidates. Figure \ref{fig:kimetalupdate} shows perihelia of all known MBCs as of 2021. Using this updated data set of MBCs and MBC candidates in the OMB, Rayleigh's z-test yields a probability $<0.001$ per cent for the longitudes of perihelion to be uniformly distributed. Comparing these asteroids to 149 946 numbered asteroids from AstDys with OMB orbits ($2.824 < a \leq 3.277\mathrm{AU}$, $e \leq 0.45$, $i \leq 40^{\circ}$), Watson's U$^2$ test returns a probability $<1$ per cent for longitudes of OMBs and those of MBCs to be drawn from the same distribution.
This updated analysis seems to confirm the clustering of MBCs suspected by \citet{kim}.

Hence, this clustering of orbital parameters could reveal an inherent property of MBCs, which would constrain their origin and evolution scenarios as a population.  Under this hypothesis, the authors predicted that MBCs were more likely to be discovered in the Northern autumn sky, when such asteroids would be observable at perihelion. \\

This article will present an imaging survey conducted to look for a tail (dust projected away from the nucleus) and/or a coma (extended atmosphere around the nucleus) surrounding targeted Main Belt Asteroids.
Section \ref{sec:obs} characterises the data products and target selection criteria. Section \ref{sec:methods} explains the reduction and analysis processes. Section \ref{sec:results} describes the results of the analysis. In sections \ref{sec:discussion} and \ref{sec:conclusion} we discusses our findings and conclusions.

%______________________________
\section{Observations}
\label{sec:obs}

Throughout three observing runs in November 2018, October 2019 and September 2020, 534 different asteroids were imaged using the Wide Field Camera (WFC) and Sloan r filter on the 2.5m Isaac Newton Telescope (INT). The full log of observations is available in Appendix \ref{appendixlog}. This section will describe the target selection criteria and the associated data products.

\subsection{Target selection}
\label{sec:targetselection}

Using the \texttt{JPL Small Body Data Base}\footnote{\href{http://ssd.jpl.nasa.gov/sbdb.cgi}{http://ssd.jpl.nasa.gov/sbdb.cgi}} (JPL SBDB) and \texttt{OpenORB} \citep{openorb}, a list of all known Main Belt asteroids meeting the selection criteria listed in Table \ref{tab:targetselection} was generated to determine target candidates. We selected asteroids:

\begin{itemize}
    \item with an apparent V-magnitude limit of less than 21 to be visible with our desired exposure times.
    \item with well determined orbits (data arc > 100 days), to be easily identified on the frames.
    \item between 50 days pre-perihelion and 200 days post-perihelion at the time of observation, to maximize chances of visible activity.
    \item with similar orbits (semi-major axis, eccentricity and Tisserand parameter) as known MBCs.
    \item with longitudes of perihelion between 0 and 30$^{\circ}$, to conform with the hypothesis of \citet{kim} (see section \ref{sec:kim}). 
\end{itemize}

660 candidates were obtained for the 2018 run, 694 for the 2019 run, and 566 for the 2020 run. During the observing runs, targets were selected in real time from the list of candidates. Priority was given to asteroids with perihelia closest to that of Jupiter, largest eccentricities, shortest time to perihelion, and belonging to young asteroid families or families containing known MBCs. The observation schedule was optimised by observing asteroids in sequences requiring small shifts of the telescope and by fitting multiple asteroids in the field of view of the WFC when possible.

\begin{figure}
    \centering
    \includegraphics[width=\linewidth]{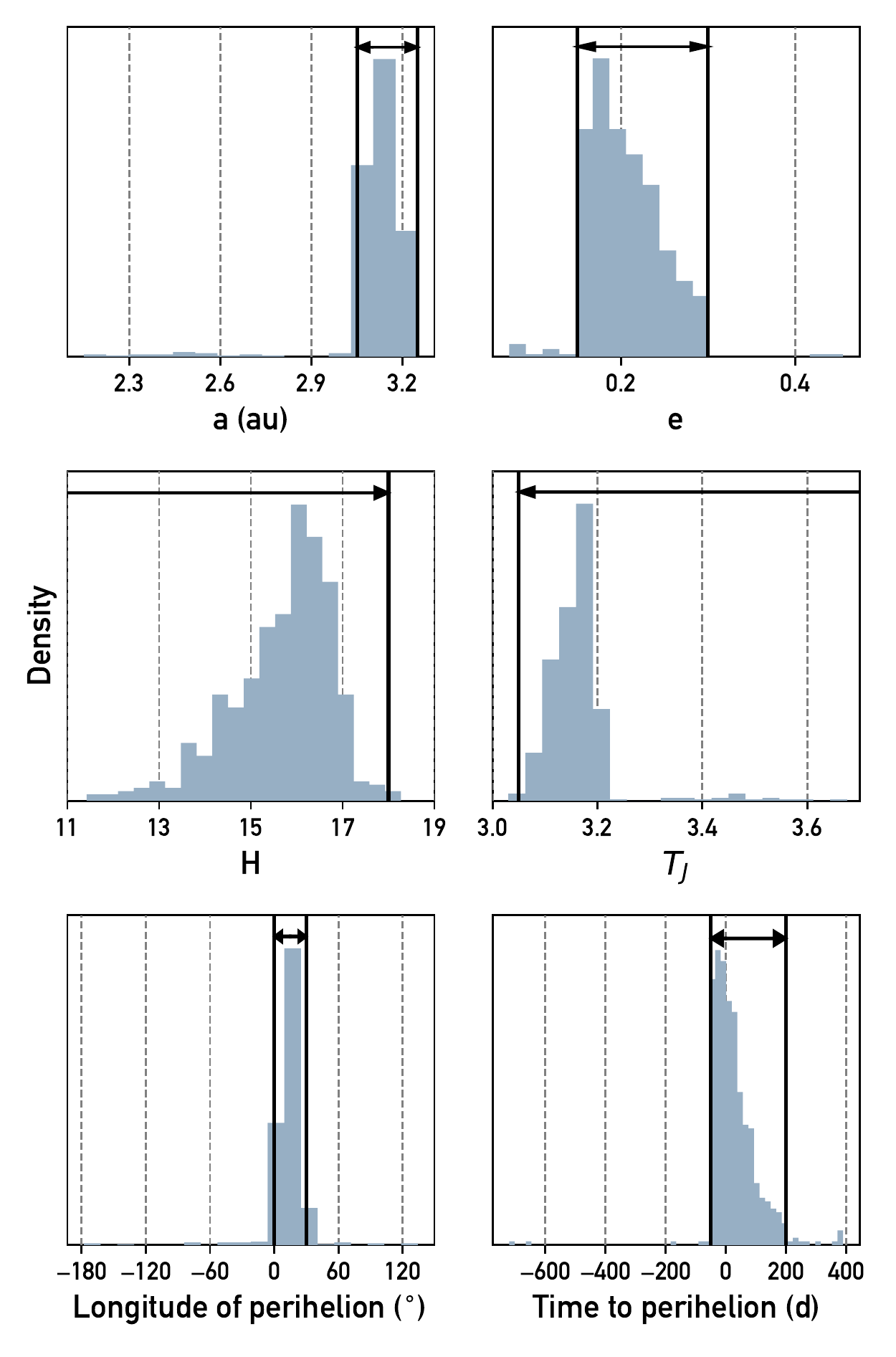}
    \caption{Distribution of orbital parameters of our 534 targets obtained from \texttt{JPL SBDB} as of 15/08/2022. Black lines and arrows indicate our selection boundaries.}
    \label{fig:orbitalparameters}
\end{figure}

Figure \ref{fig:orbitalparameters} shows orbital parameters of our targets from \texttt{JPL SBDB} as of 2022 August 15. Given the most recent calculations, while most values remain within our boundaries, a few objects do not fit our selection criteria anymore.

\begin{table*}
\centering
\footnotesize
\caption{Selection criteria for the list of potential targets. They are further described in Section \ref{sec:targetselection}.}

\begin{tabular}{lcc}
\hline
\textbf{Parameter}                                           & \textbf{Criterion}          & \textbf{Reason for the criterion}                                              \\ \hline
\textit{Absolute parameters from JPL SBDB}                   &                             &                                                                     \\ \hline
\textbf{Data arc used for ephemerides calculation}                                            & Spanning at least 100 days  & Low uncertainty on the ephemerides                                   \\
\textbf{Absolute magnitude $H$}                              & $H<18$                      & To match apparent magnitude limit of $V=21$ at time of observation \\
\textbf{Semi-major axis $a$}                                 & $3.05<a<3.25\mathrm{AU}$             & MBC-like orbit                                                      \\
\textbf{Eccentricity $e$}                                    & $0.15<e<0.3$                & MBC-like orbit                                                      \\
\textbf{Tisserand parameter with respect to Jupiter $T_{J}$} & $T_{\mathrm{J}}>3.05$                & MBC-like orbit                                                      \\
\textbf{Longitude of perihelion $\bar{\omega}$}              & $0<\bar{\omega}<30^{\circ}$ & $\bar{\omega}_{\mathrm{J}} \approx 15^{\circ}$, \citet{kim}                               \\ \hline
\textit{Time-dependent parameters calculated for the}        &                             &                                                                     \\
\textit{middle of the observing run using OpenORB  }                  &         &                                                 \\ \hline
\textbf{Closeness to perihelion}                             & $-50<\Delta t_{peri}<+200\mathrm{days}$     & Maximize chances of activity if MBC                                 \\
\textbf{Apparent magnitude $V$}                              & $V<21$                      & Target visibility                                                   \\ \hline
\end{tabular}
\label{tab:targetselection}
\end{table*}

\subsection{Data products}
\begin{table}
\centering
\caption{Number of sets of frames obtained each night, and number of sets that could be analysed by the pipeline ("valid").}
\begin{tabular}{lcc}
\hline
\textbf{Night}      & \textbf{Sets of frames} & \textbf{Valid sets of frames} \\ \hline
\textit{07/11/2018} & 73               & 40                 \\
\textit{08/11/2018} & 40               & 22                  \\
\textit{09/11/2018} & 53               & 35                  \\
\textit{10/11/2018} & 59               & 32                  \\
\textit{11/11/2018} & 19               & 10                  \\
\textit{13/11/2018} & 32               & 19                  \\ \hline
\textit{21/10/2019} & 48               & 34                  \\
\textit{22/10/2019} & 46               & 34                  \\
\textit{23/10/2019} & 54               & 28                  \\
\textit{24/10/2019} & 15               & 5                   \\ \hline
\textit{16/09/2020} & 5                & 2                   \\
\textit{17/09/2020} & 39               & 19                  \\
\textit{18/09/2020} & 42               & 21                  \\
\textit{19/09/2020} & 29               & 18                 \\ \hline
Total & 554                         & 319 \\ \hline
\end{tabular}
\label{tab:obsnights}
\end{table}

The WFC instrument has a pixel scale of $0.333$ arcsec/pixel and a total field of view of $34'\times34'$ split between 4 rectangular CCDs, each of them comprising $2048 \times 4100$ useful imaging pixels (approximately $11.4'\times22.8'$).
For each asteroid 3 to 10 frames were acquired, making up a total exposure of 500 seconds. The number of frames was selected to minimize the effects of the motion of the asteroid.
The INT has a typical seeing of 1" to 3", and each image typically shows between 500 and 2500 detectable light sources: stars, galaxies, other asteroids, etc. \\

A few asteroids were observed multiple times to double check for activity, and multiple asteroids were imaged simultaneously, which creates confusion in the vocabulary to distinguish observed objects and the associated images. In what follows, we will call a "set of frames" a single $\{$consecutive set of images$\}+\{$asteroid$\}$ combination. With this convention, two asteroids appearing on the same images count as two sets of frames, and one asteroid observed on two different nights counts as two set of frames,
On this basis, we obtained 276 sets of frames during the 2018 run, 163 during the 2019 run and 115 during the 2020 run, i.e. 554 in total (detailed in Table \ref{tab:obsnights}). The full log of observations is given in Appendix \ref{appendixlog}. 

It should be noted that the data from the 2020 run shows higher noise levels. We suspect this is due to the guiding camera having been left on during the acquisition and causing increased readout noise. Consequently, our images are not as deep as they should be, therefore the pipeline can struggle to calibrate and align some frames or to locate the faint asteroids on the images.

%______________________________
\section{Method}
\label{sec:methods}

\citet{Sonnett2011} applied activity detection algorithms to the 924 Main-Belt objects detected by TALCS but did not find any sign of activity in the data. We adapted their method to analyse our WFC data. This section will summarize the tail and coma detection techniques, and further details can be found in \citet{Sonnett2011}. 

One should note that our data products differ from the ones that the authors studied. In particular they make use of several observations of each asteroid taken at different times, and can analyse them individually to assess whether an object obtains recurring high activity detection levels. In our study almost all targets are observed only once. Therefore our study differs from theirs in the nature of what we are trying to detect as well as the characteristics of our data (seeing, precision of the data reduction, etc.).

\subsection{Data reduction and preliminary analysis}
\label{sec:reduc}

\begin{figure}
	\includegraphics[width=\columnwidth]{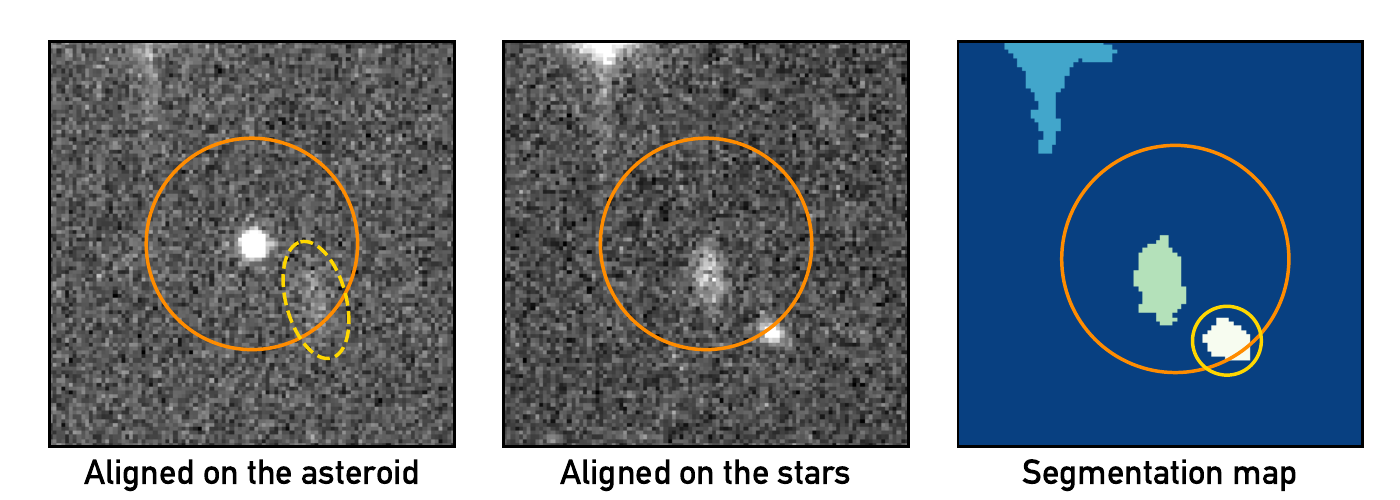}
    \caption{Left: Thumbnail of the asteroid-aligned stacked frame for asteroid 385295. The orange circle represents the $r_{\mathrm{neighbour}}$ radius within which we want no contamination by background sources. In the dashed yellow ellipse, a faint trailed background star is visible. 
    Middle: Thumbnail of the star-aligned frame. The asteroid looks trailed as it is moving between frames. The contaminating star is visible. Right: Segmentation map corresponding to the star-aligned image resulting from the source extraction, i.e. each different colour represents a different identified light source over the dark blue background. \texttt{SEP} detects flux from the contaminating star within the $r_{\mathrm{neighbour}}$ radius, allowing us to reject this set of frames.}
    \label{fig:reduc}
\end{figure}

All raw frames were bias and flat-field corrected. Defective pixels were accounted for by stacking all bias frames together and creating a mask of pixels with a significantly high level. These pixels are ignored during the analysis.

We use the \texttt{Source Extractor Python} (SEP) package \citep{sextractor,Barbary2016} to detect light sources on the images.
For each set of frames we identify matching stars and align the frames using \texttt{astroalign} \citep{Beroiz2020} before median-stacking them (see Fig. \ref{fig:reduc}, left).
Using predicted rates of motions from \texttt{JPL Horizons}\footnote{\href{https://ssd.jpl.nasa.gov/horizons/}{https://ssd.jpl.nasa.gov/horizons/}}, another stacked-image is created where the frames are aligned on the asteroid and the stars look trailed (see Fig. \ref{fig:reduc}, right).

The asteroid is identified by using World Coordinate System calibration from the \texttt{astrometry.net} \citep{Lang2010} code to look for a light source at the coordinates predicted by \texttt{JPL Horizons}. If we cannot find the asteroid this way, we use the predicted rates of motion from \texttt{JPL Horizons} to search for a light source that would be moving at similar rates on our successive frames.\\

The activity detection methods require us to see the asteroid on a uniform background, with no neighbouring star. We define a radius $r_{\mathrm{neighbour}}$ within which we must not detect any other light source, in which case we reject the set of frames (Fig. \ref{fig:reduc}). The value of $r_{\mathrm{neighbour}}$ depends on the Full Width at Half-Maximum (FWHM) of the asteroid, as described in section \ref{sec:taildetection}. 

Although we try to account for the varying noise levels and seeing, our pipeline does not perform this filtering perfectly, and sometimes fails to detect these neighbouring light sources, or accidentally mistakes background noise for a light source. To correct for this defect we manually inspect the results after running the pipeline and identify wrongly analysed targets to remove from our results, as well as wrongly rejected targets for which to force the analysis past this filtering stage. The results presented in Section \ref{sec:results} take into account this corrected filtering.

\subsection{Tail detection}
\label{sec:taildetection}

\begin{figure}
    \centering
    \includegraphics[width=\linewidth]{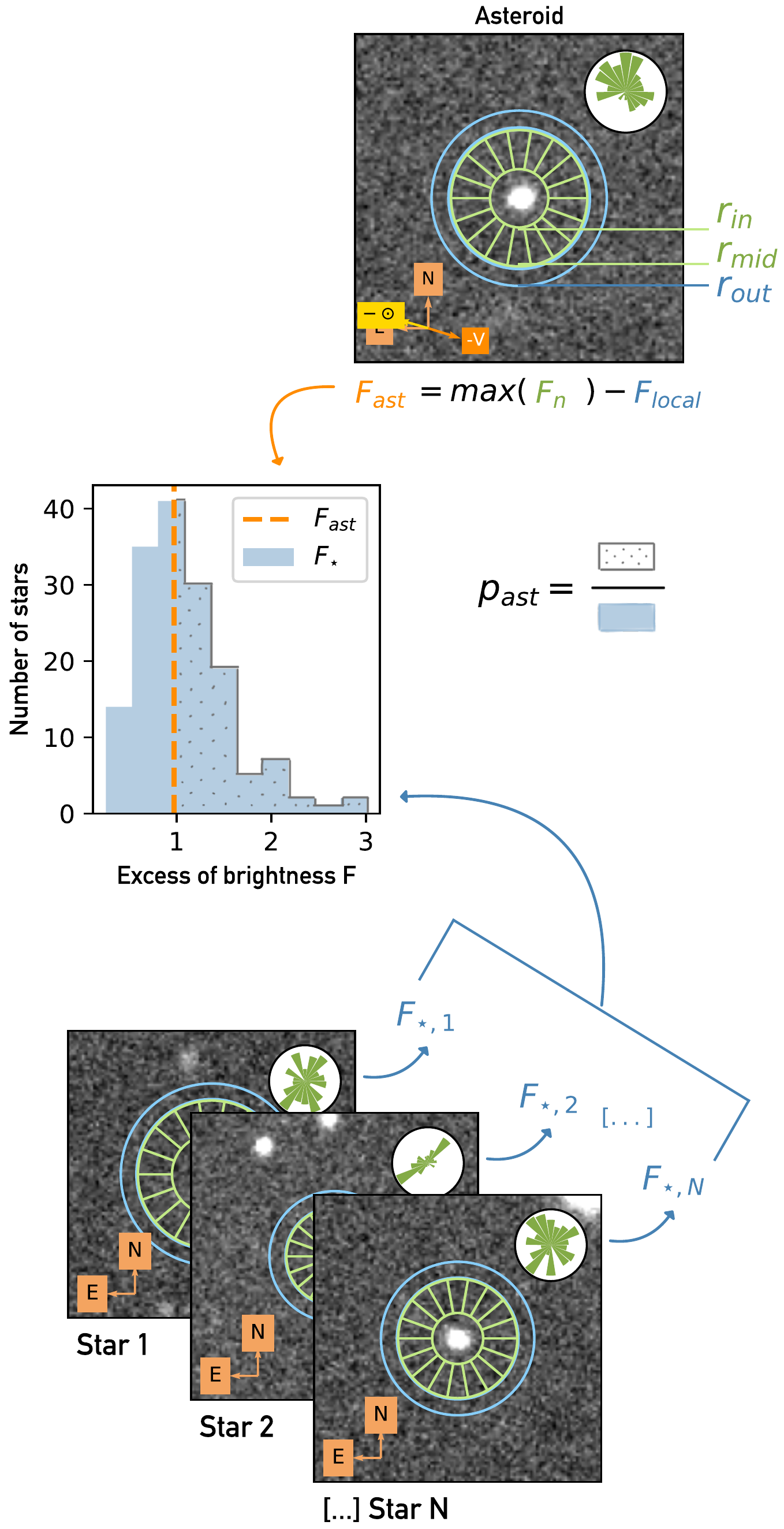}
    \caption{Principle of the tail detection analysis, as detailed in section \ref{sec:taildetection}. For the asteroid, we calculate the excess of brightness $F_{ast}$ as the difference between the median flux in the brightest green segment and the median flux of the blue annulus. Similarly we calculate values of $F_{\star}$ for a sample of comparison stars. We then calculate the tail detection indicator $p_{ast}$ as the percentage of $F_{\star}$ that are higher than $F_{ast}$. A low $p_{ast}$ value indicates that the excess of brightness in the brightest segment around the asteroid is statistically significant.} 
    \label{fig:tail}
\end{figure}

The tail detection method aims to detect an abnormally bright region adjacent to the object.
To do so we define an annulus surrounding the target and divide it into 18 angular segments. If a tail is visible we expect to detect a higher flux in the segment that it falls into (Figure \ref{fig:tail}). 

Let us call $r_{\mathrm{in}}$ and $r_{\mathrm{out}}$ the inner and outer radii of this annulus.
To account for the important seeing variations in our data, we scale $r_{\mathrm{in}}$ to the FWHM of each target. This allows us to avoid most of the flux from the nucleus, without losing too much potential tail flux. However the sensitivity of the technique depends on the area of the segments and will vary between targets. $r_{\mathrm{in}}$ is usually between 10 and 20 pixels ($\approx$ 3" to 6").
We fix the width of the annulus ($r_{\mathrm{out}} - r_{\mathrm{in}}$) at 12 pixels, or $\approx 4$ arcseconds. At 1.65AU when $r_{\mathrm{in}}=10$ pixels (seeing of $\approx1.5"$), this corresponds to a region between 4000 km to 9000 km from the nucleus.
We define a 2"-wide annulus just outside of the search annulus in order to compute the local background flux $F_{\mathrm{local}}$ as the median flux in this annulus.

For each segment n=1,2,..., 18 we compute the median flux $F_n$ in the segment in order to identify the brightest segment. We then compute the excess of brightness of this segment compared to the background, i.e. $F_{\mathrm{ast}}=max_{1\leq n \leq 18}(F_{\mathrm{n}})-F_{\mathrm{local}}$. As our observations have varying noise levels and stacking precisions, an absolute detection threshold for $F_{\mathrm{ast}}$ cannot be defined. Instead, we select stars with similar brightness to the asteroid to assess the typical noise levels and features of the frame. For each of these stars we repeat the same process and compute the excess of brightness in the brightest segment $F_{\mathrm{\star}}$ as for the asteroid. We then compare $F_{\mathrm{ast}}$ to the distribution of these $F_{\star}$ values to see whether the asteroid stands out. To do so, following the methodology of \citet{Sonnett2011} we compute a tail detection indicator $p_{\mathrm{ast}}$ corresponding to the percentage of stars that have $F_{\star}>F_{\mathrm{ast}}$. For instance, $p_{\mathrm{ast}}=0.08$ means that only $8$ per cent of the stars had $F_{\star}>F_{\mathrm{ast}}$. If there is a tail, $F_{\mathrm{ast}}$ should be significantly larger than most values of $F_{\star}$ therefore $p_{\mathrm{ast}}$ should be low. However, without a tail the asteroid should look similar to the stars and $F_{\mathrm{ast}}$ could randomly rank anywhere among the $F_{\star}$ values, therefore a low $p_{\mathrm{ast}}$ does not guarantee a detection. 

To counter this statistical variability, \citet{Sonnett2011} use a large number of frames per target. They compute $p_{\mathrm{ast}}$ for each individual frame and analyse the distribution of $p_{\mathrm{ast}}$ values, looking for consistently small values.
In our study we only usually analyse one stacked image per asteroid, so our one value of $p_{\mathrm{ast}}$ is our final indicator to assess which objects are worth examining manually. Detection thresholds will be discussed in section \ref{sec:results}. 

\subsection{Coma detection}
\label{sec:methodcoma}

The coma detection method compares the asteroid’s brightness profile to the Point Spread Function (PSF) expected for a non-extended body. As our exposure times were selected to minimize the effects of the asteroid’s motion, it should have the same PSF as stars. 

On each frame of the set we extract flux-normalised thumbnails of stars that are slightly brighter than the asteroid and in the same region on the CCD. We want similar stars that have a good signal to noise ratio in order to build a good quality PSF model, and we want to avoid distortion effects across the CCD. We only perform the analysis if at least 2 comparison stars are found. As we are going to median-stack the images, a high number of comparison stars is preferable to minimize the effects of the noise or of visual anomalies present on individual thumbnails of the stars. A low number of comparison stars found to be similar to the asteroid can also indicate a problem with the asteroid thumbnail (undetected close stars, visual artefact, wrong object mistaken for the asteroid, etc.).

The following process is illustrated by Fig. 6 in \citet{Sonnett2011}. The thumbnails of the stars are median-stacked to obtain a PSF specific to each frame. 
Each PSF thumbnail is then artificially trailed with the rates of motion of the asteroid to create a frame-specific simulated asteroid profile.
The resulting frame-specific thumbnails are then stacked together to create the final simulated asteroid model.

A comet model is also generated by convolving a $1/r$ profile with the asteroid model. It has a more diffuse profile for the same total brightness.

Following the method of \citet{Sonnett2011}, we compare these models to the asteroid by fitting a linear mix of the two flux-normalised models [$(1-f_{\mathrm{c}}) \times \mathrm{asteroid\,model} + f_{c} \times \mathrm{coma\,model}$] to the flux-normalised asteroid thumbnail to determine the coma fraction $f_c$. If the asteroid displays a coma, we can expect a more extended profile and therefore a higher value of $f_c$ than for a non-active body in these given observing conditions. We assess what constitute a "high" value of $f_c$ by using a control sample of stars, and discuss detection thresholds in section \ref{sec:results}.

\subsection{Validation}
\label{sec:testdata}

In order to test the ability of the analysis pipeline to detect activity and better constrain what constitute low or high detection levels, we performed the analysis on images of a {known} active object observed with the WFC.

\begin{figure*}
    \centering
    \includegraphics[width=\linewidth]{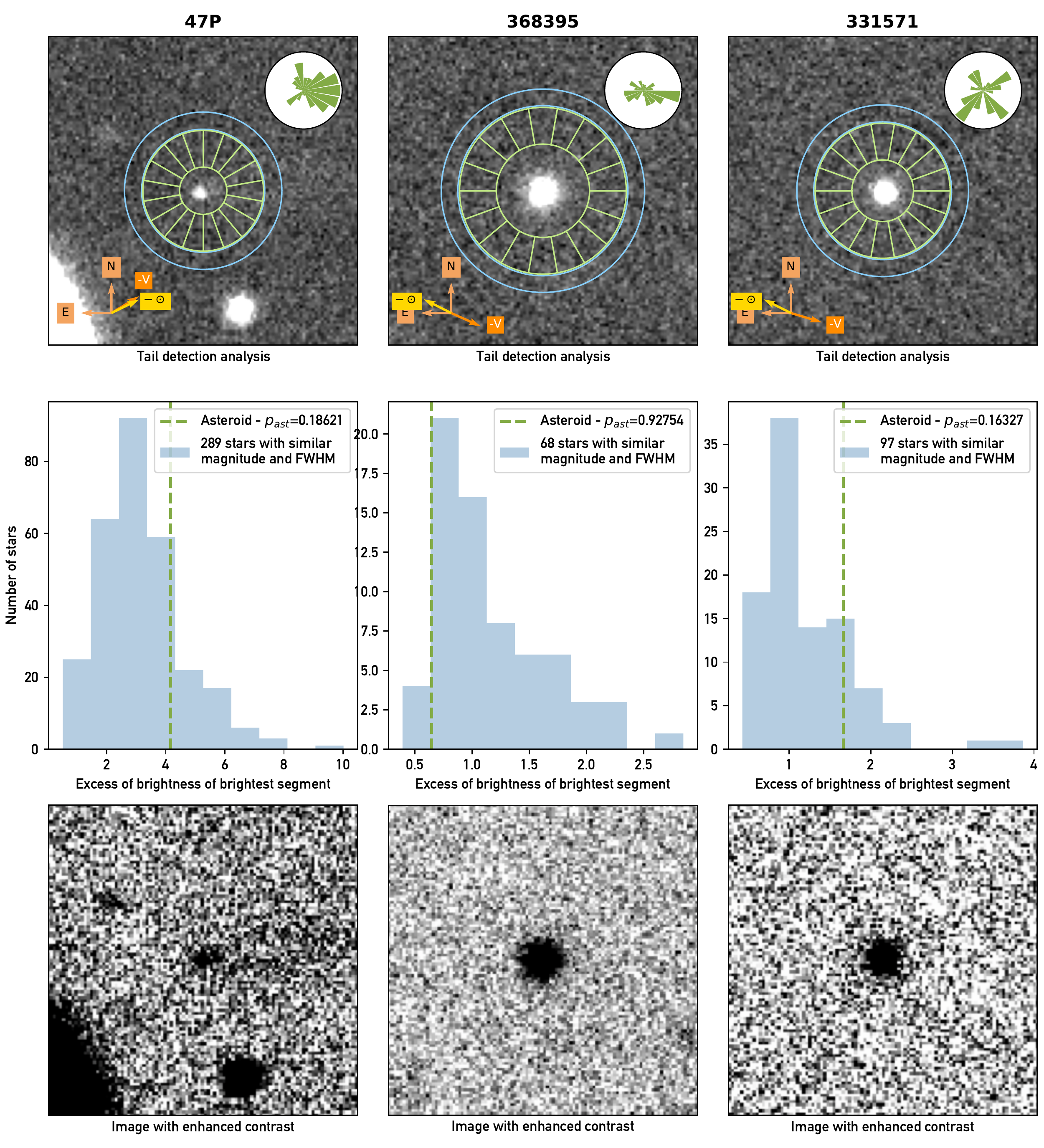}
    \caption{Result of the analysis of Jupiter Family Comet 47P, as well as two of our asteroids that appear to be inactive for reference. 47P shows a low $p_{\mathrm{ast}}$ value (middle left figure), with brightness concentrated in the antisolar/antivelocity direction (top left figure). When displaying the image with a more appropriate brightness scale (bottom left) a faint tail is visible in this direction. Asteroid 368395 also shows brightness in one particular direction (top middle figure) but it corresponds to a high $p_{\mathrm{ast}}$ value (central figure) which indicates that this level of brightness is not significant. Finally asteroid 331571 obtains a low $p_{\mathrm{ast}}$ value (middle right figure) but multiple segments around the asteroids obtain similarly high levels of brightness, there is no tail-like feature that stands out.}
    \label{fig:47Pandrandom}
\end{figure*}

47P/Ashbrook–Jackson is a Jupiter-family comet. It was observed on 2006 March 02 with the INT (Fig. \ref{fig:47Pandrandom}) \citep{snodgrass2007} {while at 5.1 au from the Sun, 35 months pre-perihelion. Therefore it is only displaying faint activity, which matches the level of activity we typically aim to detect around our targets. At first glance one might not notice activity when reviewing the image of 47P, however an appropriate brightness scale reveals a faint tail in the West direction.}
The pipeline returned a $p_{\mathrm{ast}}$ value of 0.186 with the brightest segment matching the antisolar direction. $p_{\mathrm{ast}}$ is relatively low but not minimal, as the brightness seems divided between three main segments. 47P also obtained a coma fraction of 0.046.

For comparison, Figure \ref{fig:47Pandrandom} shows the result of the analysis for two random asteroid that do not seem active.  Asteroid 368395 shows brightness in one particular direction but it does not physically match the direction of a potential tail, moreover it obtains a $p_{\mathrm{ast}}$ value of 0.93, indicating that the excess of brightness of the bright direction is not statistically significant. Asteroid 331571 obtained a $p_{\mathrm{ast}}$ value of 0.16327, but multiple directions obtain similar levels of high brightness, indicating that the low $p_{\mathrm{ast}}$ value is not due to a tail. Both of these objects obtained a coma fraction of 0.0001, significantly lower than 47P. 

%______________________________
\section{Results}
\label{sec:results}

\subsection{Overview}

\begin{figure*}
	\includegraphics[width=0.99\linewidth]{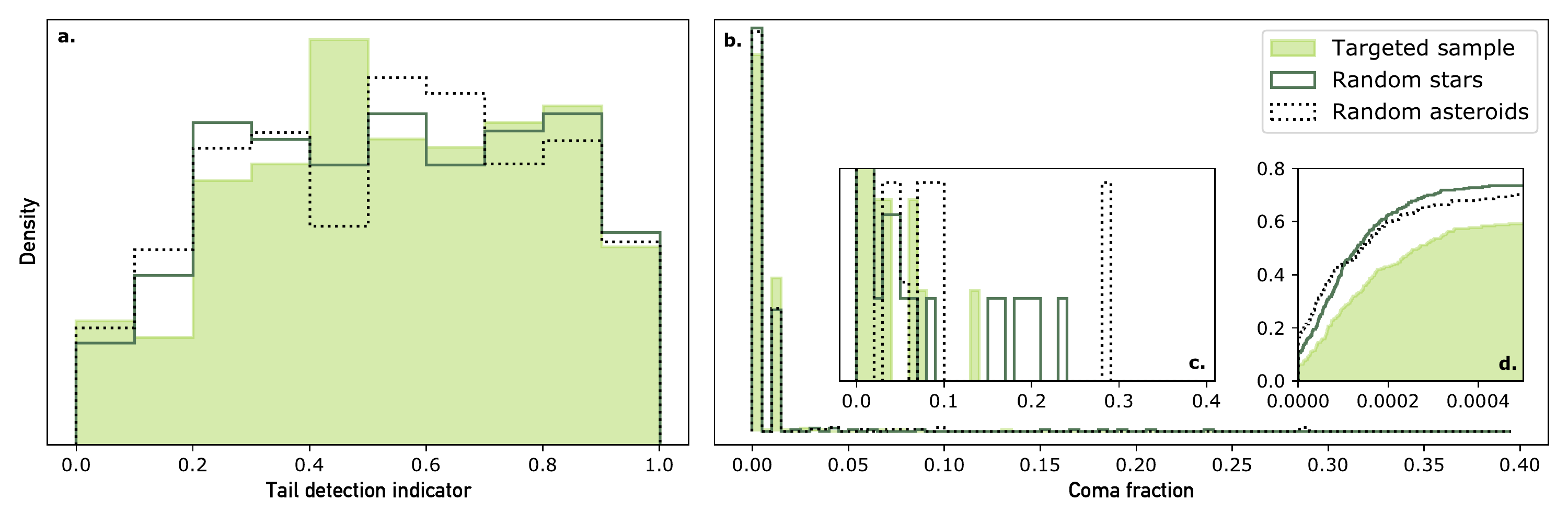}
    \caption{Normalized histograms of the $p_{\mathrm{ast}}$ values (a) and coma fractions (b) obtained for our targeted sample and control samples of random stars and asteroids. Figure (c) is a zoomed in version of Figure (b) highlighting the few values between 0.02 and 0.4. Figure (d) shows the cumulative distributions of the values within 0 and 0.0005.}
    \label{fig:rank_histogram}
\end{figure*}

Out of 549 sets of frames, we obtained valid $p_{\mathrm{ast}}$ values for 319 sets, corresponding to 316 distinct asteroids, and valid coma fractions for 293 of them (as the selection of comparison stars used for the coma detection method is more strict than for the tail detection method).  %(XXX might need to add 'forced' to that) (XXX 380 / 289 unfiltered) 
Most of the rejections are due to close neighbouring sources, but in a few cases the asteroids were too faint to be found (especially for data from the 2020 run).
The sets of frames for which the analysis could not be performed were visually checked for signs of activity.

The distribution of tail and coma detection scores obtained among our targeted sample is shown by the filled histogram on Figure \ref{fig:rank_histogram}, and will be commented on in Sections \ref{sec:tailresults} and \ref{sec:comaresults}.

For each method, we determined arbitrary detection thresholds from: 1) Studying the total distribution of results among our sample. 2) Taking into account our test results for 47P.  3) Applying the pipeline to stars to evaluate what results are expected from non detection. 4) Applying our pipeline to random asteroids that appear on our frames coincidentally. 
These detection thresholds do not guarantee a detection but determine which asteroid should be further investigated manually for activity. This will be detailed progressively in the next subsections.

\subsection{Tail detection results}
\label{sec:tailresults}

The distribution of $p_{\mathrm{ast}}$ values obtained among our targeted sample and control samples is shown in Figure \ref{fig:rank_histogram} (a). 
We obtain a slightly sloped distribution favouring medium/high values of $p_{\mathrm{ast}}$.

In the absence of activity, one would expect a uniform distribution, as each asteroid is equivalent to its comparison stars and can therefore reach any value of $p_{\mathrm{ast}}$ randomly. 
In reality, the overall shape of the distribution is affected by the pipeline's inability to perfectly detect neighbouring stars. 

For instance, on Figure \ref{fig:47Pandrandom} (middle row right), the distribution of $F_{\star}$ shows most values concentrating around 1, with a second small population around 3.5.  This is likely due to neighbouring sources close to these stars going undetected by the pipeline. Although the asteroids around which a neighbouring star went undetected can be removed manually from the results, for a given asteroid we do not remove the comparison stars that have an undetected close neighbour. These stars will obtain high $F_{star}$ values, likely higher than $F_{ast}$, biasing $p_{\mathrm{ast}}$ towards higher values. Detection thresholds take this bias into account and will be discussed in section \ref{sec:highdetectionscores}.

\subsection{Coma detection results}
\label{sec:comaresults}

The distribution of coma detection scores obtained among our targeted sample is shown in Figure \ref{fig:rank_histogram} (b). 
Most asteroids obtain coma fractions lower than 0.05 per cent, with a second peak at 1 per cent due to it being the initial condition for the fitting process, and a few values spreading out to 2.9 per cent. 

The initial value of 0.01 was selected in order not to encourage the exploration of higher coma fractions and not bias it towards small values. There does not seem to be any particular feature on the images explaining the fitting process stagnating around 0.01.

\subsection{Control Samples}
\label{sec:controlsamples}

To control the behaviour of the analysis pipeline we applied it to two control samples comprised of non-active sources. 
We first applied the pipeline to a first control sample of 312 stars appearing on multiple frames. The distribution of scores for these stars is shown with  as the solid line in Figure \ref{fig:rank_histogram}.
We also applied the pipeline to a sample of 339 non-targeted asteroids. These are asteroids that randomly happen to be in the field of view of the instrument during the observations. We used the \texttt{Skybot} \citep{skybot} service to list these objects. The distribution of scores for these random asteroids is shown with a dotted outline in Figure \ref{fig:rank_histogram}.

We used Kolmogorov-Smirnov statistics to compare the distribution of results for our targets and our control samples. When comparing two data samples, if the p-value returned by the KS test is lower than 5 per cent we can reject the hypothesis that the samples are drawn from the same distribution. Table \ref{tab:kolmogorov} shows the p-values derived from the KS test.

\begin{figure}
    \centering
    \includegraphics[width=\linewidth]{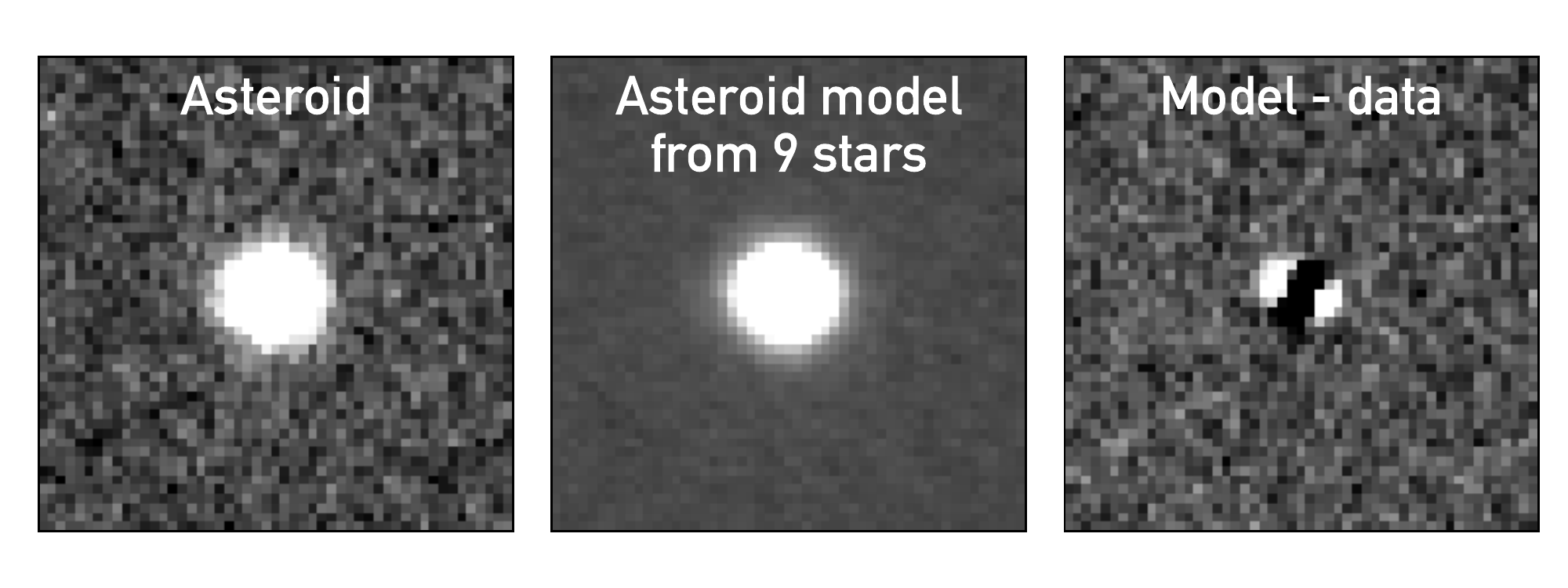}
    \caption{Example of a case where the asteroid model does not accurately replicate the motion of the asteroid. The "model - data" image on the right is the result of subtracting the asteroid thumbnail (left) to the asteroid model (middle). The model seems slightly more trailed than the asteroid. }
    \label{fig:comatrail}
\end{figure}

For the tail detection, the KS test returns high p-values, indicating that the distributions are similar for the targets and both control samples. However, the KS test indicates differences in the distributions of coma fraction.  Fig \ref{fig:rank_histogram} d. shows the cumulative distribution of the results for all three samples close to zero, in the range where most of the values are found. The targeted sample and the random asteroids show similar slopes whereas the sample of stars shows a steeper slope. This indicates a different behaviour when analysing  moving objects compared to fixed sources. 
Although we try to replicate the motion of the asteroids when building the asteroid models, the trailing is not perfect and some residuals remain, as illustrated by Figure \ref{fig:comatrail}.

One should also note that the random asteroids were studied on frames from a few nights only, and the stars from one night only. Although we tried to select frames and nights that were representative of the variations in seeing in our total sample, this can explain the differences in the distribution of results, and show how sensitive the coma method is to the quality of the data. \\

\begin{table}
\centering
\caption{p-values from the Kolmogorov-Smirnov test used to compare the distributions of results between our targets and each control sample. "R-Ast" = Random asteroids.}
\begin{tabular}{lccc}
\cline{2-4}
     & Targets vs. Stars & Targets vs. R-Ast. & Stars vs R-Ast. \\ \hline
Tail & p=0.67            & p=  0.30     & p= 0.97                    \\ \hline
Coma & p=5.0$\times10^{-6}$           & p=5.9$\times10^{-5}$     & p=0.12                     \\ \hline
\end{tabular}
\label{tab:kolmogorov}
\end{table}

\subsection{Manual inspection of targets with high detection scores}
\label{sec:highdetectionscores}

Given the previous results, we chose to manually review the objects with $p_{\mathrm{ast}}\leq0.31$, corresponding approximately to the top 20 per cent of objects with lowest $p_{\mathrm{ast}}$ values. Given the artefacts in the coma detection method previously mentioned and how sensitive it is to the quality of the images and to the selection of reference stars, we decided not to use it as our main detection tool. Still, the result of the coma detection method can be a useful indicator to support detection in the case of a high tail detection probability, and the asteroid model can be subtracted from the image to try to isolate and inspect any tail-like feature.

For each object with $p_{\mathrm{ast}}\leq0.31$ we look for a visible coma or a visible tail. The features that we are looking for are: 

\begin{itemize}
    \item the brightest segment(s) being significantly brighter than the segments in other directions. If multiple directions show similar levels of brightness as the brightest direction then it does not really stand out as a "tail".
    \item the direction of the brightest segment being close to the antivelocity/antisolar directions.
    \item a visible tail-looking feature. 
\end{itemize} 

Without most of these conditions it is likely that the high tail detection level is simply random or due to background contamination, and we need to investigate no further.

\begin{figure}
    \centering
    \includegraphics[width=\linewidth]{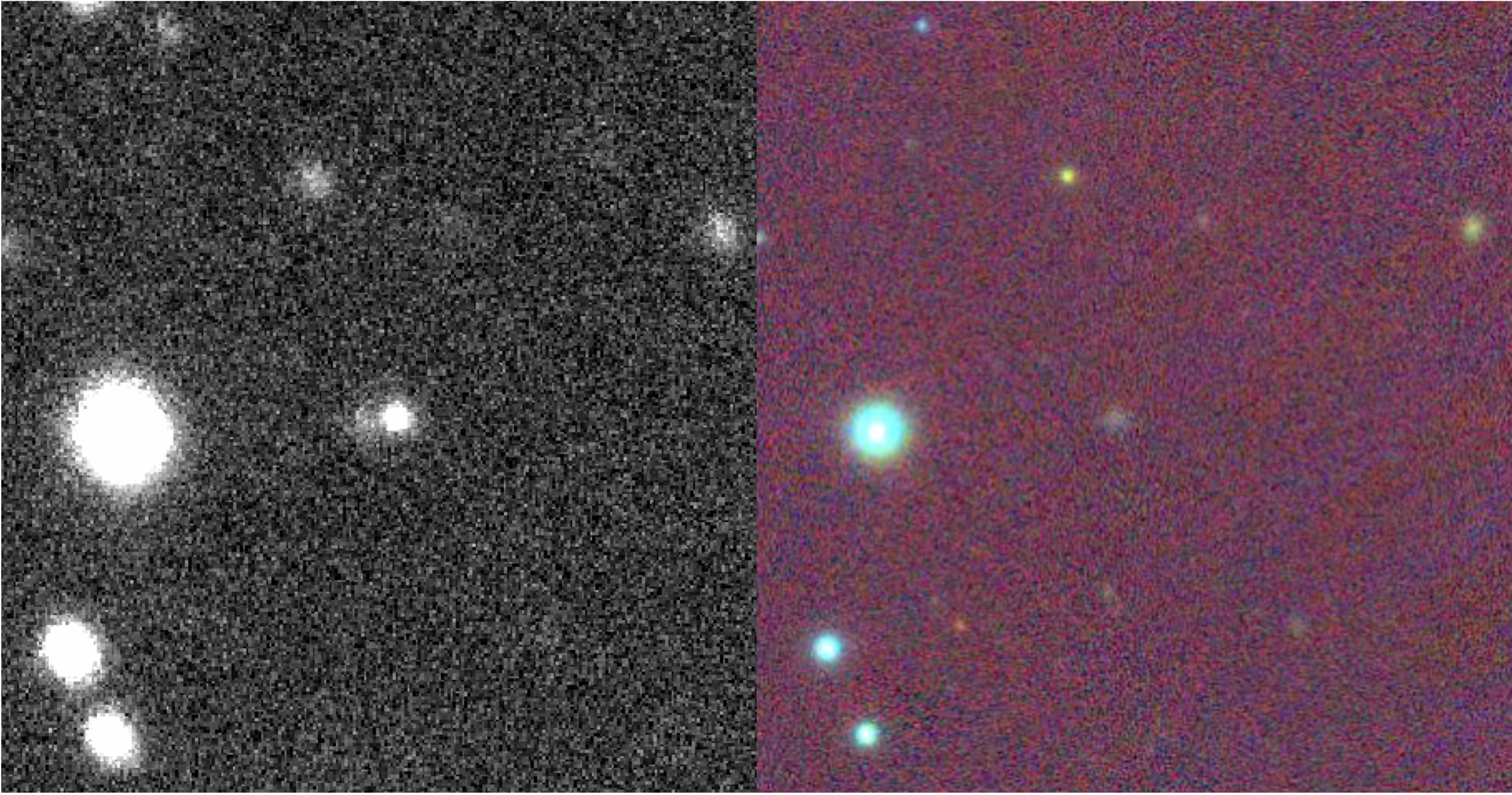}
    \caption{Example of a false detection due to a background source. On our WFC image (left), asteroid 369238 shows a tail like feature on our image. In reality it is a faint background source, visible on the Panstarrs-1 background image (right). }
    \label{fig:PS1falsedetection}
\end{figure}

For asteroids satisfying some of these conditions, we also downloaded images of the background fields from the \texttt{PanSTARRS DR1} (PS1) \texttt{Image Cutout Server}\footnote{\href{https://ps1images.stsci.edu/cgi-bin/ps1cutouts}{https://ps1images.stsci.edu/cgi-bin/ps1cutouts}} to see whether these detection levels can be due to a faint background source close to the asteroid that we would not have been able to discern on the image. However, our observations are typically slightly deeper than the PS1 data and a background source could still go unnoticed.
Fig. \ref{fig:PS1falsedetection} shows an example of an asteroid that obtained high detection levels and showed a comet-like feature that turned out to be a faint background star detected via PS1.

Tables \ref{tab:highest} and \ref{tab:highest2} lists our results and comments for these asteroids. In most cases the detection turned out to be due to faint background sources. In many other cases, although no background source was visible, no tail-like features were visible either, and it seems like the asteroids reached low $p_{\mathrm{ast}}$ values randomly. However, although not unambiguously displaying a tail, asteroid 2001 NL19 (279870) caught our attention. It is  described in the following subsection.

\begin{table*}
\begin{tabular}{lcccccccc} \hline
Night & Object     & $p_{\mathrm{ast}}$ & $f_{\mathrm{c}}$      & RA \& DEC              & One bright dir. & AS/AV & Tail-Like & Bkg. source                            \\ \hline
07-11-2018     & 2001 RB105 & 0.1  & 2.0e-05    & 22:37:42.230 -04:45:14.30 & N                         & AS                      & N                  & N                                          \\
07-11-2018    & 2007 SH20  & 0.31 & 8.4e-11    & 22:42:46.594 -16:52:19.33 & Y                         & $\sim$AS                & Y                  & N                                          \\
07-11-2018     & 82779      & 0.28 & 0.02        & 03:13:2.870 +15:09:3.96   & N                         & N                       & N                  & N                                          \\
07-11-2018     & 211018     & 0.28 & 0.01        & 03:48 21.134 +18:07:38.21 & N                         & N                       & ?                  & ?                                          \\
07-11-2018     & 279870     & 0.15 & 0.01 & 05:15:34.286 +29:14:59.09 & Y                         & AV                      & Y                  & N                                          \\
07-11-2018     & 466499     & 0.1  & 1.5e-04    & 00:16:28.664 -24:02:53.53 & Y                         & N                       & ?                  & Y                                          \\ \hline
08-11-2018     & 163644     & 0.24 & 0.01        &  23:53:18.67 +05:04:03.5          & N                         & N                       & N                  & N                                          \\
08-11-2018     & 27543      & 0.2  & 0.01        & 23:26:14.496 -14:50:49.34 & Y                         & AV                      & Y                  & Y                                          \\
08-11-2018     & 336023     & 0.24 & 2.3E-04 & 00:57:13.977 -03:14:33.68 & Y                         & N                       & N                  & N                                          \\
08-11-2018     & 282183     & 0.28 & 3.1E-04 & 00:45:23.337 +06:57:10.58 & Y                         & N                       & ?                  & N                                          \\
08-11-2018     & 7279       & 0.31 & N/A          &   01:35:12.31 +13:18:32.2        & Y                & AS                      & N                  &                 N                           \\
08-11-2018     & 5575       & 0.31 & N/A          &    01:26:21.71 +08:06:42.6              & N                 & AV                      & N                  &       N                                     \\
08-11-2018     & 113191     & 0.3  & 9.9E-04      &   02:17:27.18 +15:20:27.9      & N                         & AS                      & N                  &         N                                   \\ \hline
09-11-2018    & 206129     & 0.29 & 1.2E-04 & 23:13:31.998 -06:48:19.34 & Y                         & N                       & N                  & Y                                          \\
09-11-2018      & 280260     & 0.03 & 4.2E-05    &  02:16:46.89 +15:18:47.0 & Y                         & N                       & Y                  &   N                                         \\
09-11-2018      & 2001 UF200 & 0.03 & 0.034 & 00:05:09.87 +01:26:21.6  & Y                         & N                       & Y                  & Y                                          \\
09-11-2018      & 337928     & 0.09 & N/A          & 00:55:38.939 +22:11:19.34 & Y                         & N                       & Y                  & Y                                          \\ \hline
10-11-2018      & 105073     & 0.17 & 5.94E-05    & 22:34:44.628 -10:14:8.49  & Y                         & AV                      & Y                  & Y                                          \\
10-11-2018     & 331571     & 0.16 & 1.6E-04 &  22:54:00.30 -18:24:23.4 & N                         & N                       & N                  &        N                                    \\
10-11-2018     & 2007 TH5   & 0.22 & 1.9E-04 & 01:54:23.764 +13:08:30.24   & Y                         & N                       &      ~N              &       ~N                                     \\
10-11-2018     & 382502     & 0.13 & 6E-04  &  03:31:50.35 +14:54:27.5 & N                         & N                       & N                  & Y                                          \\
10-11-2018     & 424240     & 0.28 & 1.6E-04 &  00:55:50.49 -00:27:38.8  & N                         & N                       & N                  &    N                                        \\ \hline
11-11-2018     & 203846     & 0.07 & 2.1E-04   &  23:30:00.59 +01:30:28.1 & Y                         & N                       & Y?                 & Y                                          \\ \hline
13-11-2018     & 203861     & 0.27 & 6.6E-05   & 00:44:27.639 -04:29:40.38 & Y                         & N                       & Y?                 & Y                                          \\
13-11-2018     & 161086     & 0.1  & 0.01        & 00:07:28.388 +08:02:39.18 & Y                         & N                       & Y?                 & Y                                          \\
13-11-2018     & 7726       & 0.22 & 0.01        & 00:53:37.367 +13:34:36.00 & Y                         & $\sim$AV                & N                  & N                                          \\ \hline
21-10-2019     & 117595     & 0.13 & 1.4E-04 & 22:32:33.401 +00:01:14.30 & Y$\sim$                   & N                       & Y                  & Y                                          \\
21-10-2019      & 369238     & 0.11 & 0.03 & 23:51:46.307 +12:59:28.80 & Y                         & N                       & Y                  & Y                                          \\
21-10-2019      & 232276     & 0.19 & 3.9E-05    & 01:05:8.024 +04:14:5.91   & N                         & N                       & $\sim$             & $\sim$                                     \\
21-10-2019      & 2008 UF56  & 0.21 & 9.6E-10   & 02:03:8.414 +15:00:39.34  & Y                         & N                       & N                  & Y                                          \\
21-10-2019      & 2013 RD4   & 0.2  & 2.03E-05    &  03:24:24.25 +15:27:47.1   & Y$\sim$                   & N                       & N                  &  N                                          \\ \hline
22-10-2019      & 460700     & 0.09 & 1E-09    &  22:27:25.28 +14:50:29.5  & Y                         & N                       & N                  & BP            \\
22-10-2019     & 191403     & 0.07 & 7.5E-03 & 22:45:43.024 -01:02:45.46 & N                         & N$\sim$                 &        N            & Y                                          \\
22-10-2019     & 366484     & 0.22 & 2.7E-04  & 22:54:3.732 +16:50:54.09  & Y                         & N                       & N                  & BP \\
22-10-2019     & 8339       & 0.09 & 0.01        & 23:44:10.759 -01:32:58.68 & Y                         & N                       & N                  & N                                          \\
22-10-2019     & 44009      & 0.2  & 0.01        & 01:19:6.757 +01:12:22.45  & Y                         & N                       & $\sim$             & Y                                          \\
22-10-2019     & 410664     & 0    & 0.01        & 01:12:33.445 +04:50:46.09 & Y                         & N                       & Y                  & Y                                          \\
22-10-2019     & 2008 TB65  & 0.31 & 3.5E-04 & 01:33:22.90 +22:20:46.2  & Y                         & N                       & N                  &   N                                         \\
22-10-2019     & 2008 UH353 & 0.27 & 1.5E-04 & 01:55:39.36 +19:28:42.1  & $\sim$N                   & N                       & N                  &   N                                         \\
22-10-2019     & 4592       & 0.25 & N/A         & 02:24:36.62 +13:53:21.2  & Y                         & N                       & N                  & N                                          \\
22-10-2019     & 15657      & 0.21 & 0.01        &  04:10:09.32 +20:03:24.4 & N                         & N                       & N                  &  N                                          \\
22-10-2019     & 274485     & 0.21 & 1.2E-04 &  02:24:44.37 +13:43:03.8    & Y                         & N                       & N                  & BP                    \\ \hline
23-10-2019     & 157853     & 0.27 & 6.4E-05    &  22:37:38.83 -12:15:46.7  & Y                         & N                       & N                  &    N                                        \\
23-10-2019     & 2014 WR33  & 0.29 & 7.4E-05    &  23:23:05.76 -14:06:01.4  & N                         & N                       & N                  &    N                                        \\
23-10-2019     & 197174     & 0.22 & 2.2E-04 & 23:37:37.344 -11:14:35.66 & $\sim$Y                   & AV                      & $\sim$Y            & Y (very faint)                               \\
23-10-2019     & 195686     & 0.08 & 0.01        & 00:45:27.282 +02:14:24.86 & $\sim$N                   & $\sim$AV                & N                  & Y (very faint)                               \\
23-10-2019     & 233649     & 0.22 & 0.01        & 01:14:4.052 +33:33:39.52  & N                         & $\sim$AV                & N                  & N                                          \\
23-10-2019     & 332561     & 0.14 & 9.2E-05    & 03:17:19.058 +13:45:8.45  & $\sim$N                   & N                       & N                  & N                                          \\ \hline

\end{tabular}
\caption{
Targets that obtained the highest tail detection results and comments based on our manual review ("Y"= Yes / "N" = No). For each object, we are interested in whether only one bright direction is standing out ("One bright dir."), whether this direction matches the antisolar or antivelocity directions ("AS/AV"), whether a tail-like feature is visible by eye ("Tail-like"), and whether we can see a backgroung source on PanSTARRS-1 images ("Bkg source"). "BP" = There seems to be one particularly bright pixel that might cause the high detection level. The coma fraction is "N/A" when not enough comparison stars were found to perform the coma detection analysis. (Part 1/2)}
\label{tab:highest}
\end{table*}

\begin{table*}
\begin{tabular}{lcccccccc} \hline
Night & Object     & $p_{\mathrm{ast}}$ & Coma fraction        & RA \& DEC               & One bright dir. & AS/AV & Tail-Like & Bkg. source                           \\ \hline
17-09-2020 & 451133 & 0.08 & 8.7E-10 & 23:08:58.92 +07:09:50.2 & Y & $\sim$AV & $\sim$ & $\sim$Y (very faint) \\
17-09-2020    & 220871     & 0.07    & 2.3E-04          & 00:41:55.463 +01:38:31.90 & Y                         & N                       & $\sim$Y            & N                                          \\
17-09-2020    & 276550     & 0.24    & 9.6E-09          & 00:58:18.12 +05:13:36.4 & $\sim$N                   & N                       & N                  & N                                           \\
17-09-2020  & 2015 XX73 & 0.22 & 1.3E-04 & 01:06:07.46 +16:39:47.0 & N & N & N & N \\
17-09-2020 & 2009 SK199 & 0.19 & 6.7E-10 & 01:08:06.64 +13:13:34.0 & $\sim$Y & $\sim$AS & N & $\sim$Y (very faint)  \\
17-09-2020    & 218444     & 0.2  & 0.01        & 01:08:25.646 +19:27:9.82  & $\sim$Y                   & N                       & Y                  &         N                                   \\ 
17-09-2020 & 67214 & 0 & 0.01 & 01:53:22.21 +10:51:38.7 & Y & N & N & Y (faint) \\\hline
18-09-2020 & 2014 OF367 & 0.2 & 9.6E-09 & 00:35:26.97 +26:32:46.6 & Y & N & N & N \\
18-09-2020    & 330753     & 0.26 & 1.7E-04 & 04:28:23.706 +06:19:38.22 & Y                         & N                       &         N           & N                                          \\
18-09-2020    & 223356     & 0.08 & N/A          & 01:45:05.92 +08:06:22.1    & Y                         & N                       & N                  &    N                                        \\ \hline
19-09-2020    & 487327 & 0.3 & 2.6E-04 & 22:52:33.77 -02:30:36.9 & Y &$\sim$AS & N & N  \\
19-09-2020    & 2009 QN24 & 0.22 & 2.5E-04 & 23:09:40.80 -03:34:20.9  & Y & N & N & Y \\
19-09-2020    & 260356 & 0.21 & 2.4E-04 & 00:59:27.25 +19:46:48.4  & N & $\sim$AV & N & N \\
19-09-2020    & 133979 & 0.23 & N/A &  23:56:54.58 -16:14:29.9  & Y & N & N & N \\
19-09-2020    & 452265     & 0.16    & 6.6E-05          &   23:50:49.78 -16:29:49.4                        & Y                         & N                       & N                  &      N                                      \\
19-09-2020    & 2014 OC169 & 0.25    & 7.4E-05        &  01:23:54.46 +14:11:02.1   & N                         & $\sim$AV                & N                  &  N                                         \\ \hline
\end{tabular}
\caption{
Targets that obtained the highest tail detection results and comments based on our manual review ("Y"= Yes / "N" = No). For each object, we are interested in whether only one bright direction is standing out ("One bright dir."), whether this direction matches the antisolar or antivelocity directions ("AS/AV"), whether a tail-like feature is visible by eye ("Tail-like"), and whether we can see a backgroung source on PanSTARRS-1 images ("Bkg source"). "BP" = There seems to be one particularly bright pixel that might cause the high detection level. The coma fraction is "N/A" when not enough comparison stars were found to perform the coma detection analysis. (Part 2/2)}
\label{tab:highest2}
\end{table*}

\subsection{2001 NL19 (279870)}
\begin{figure*}
    \centering
    \includegraphics[width=\linewidth]{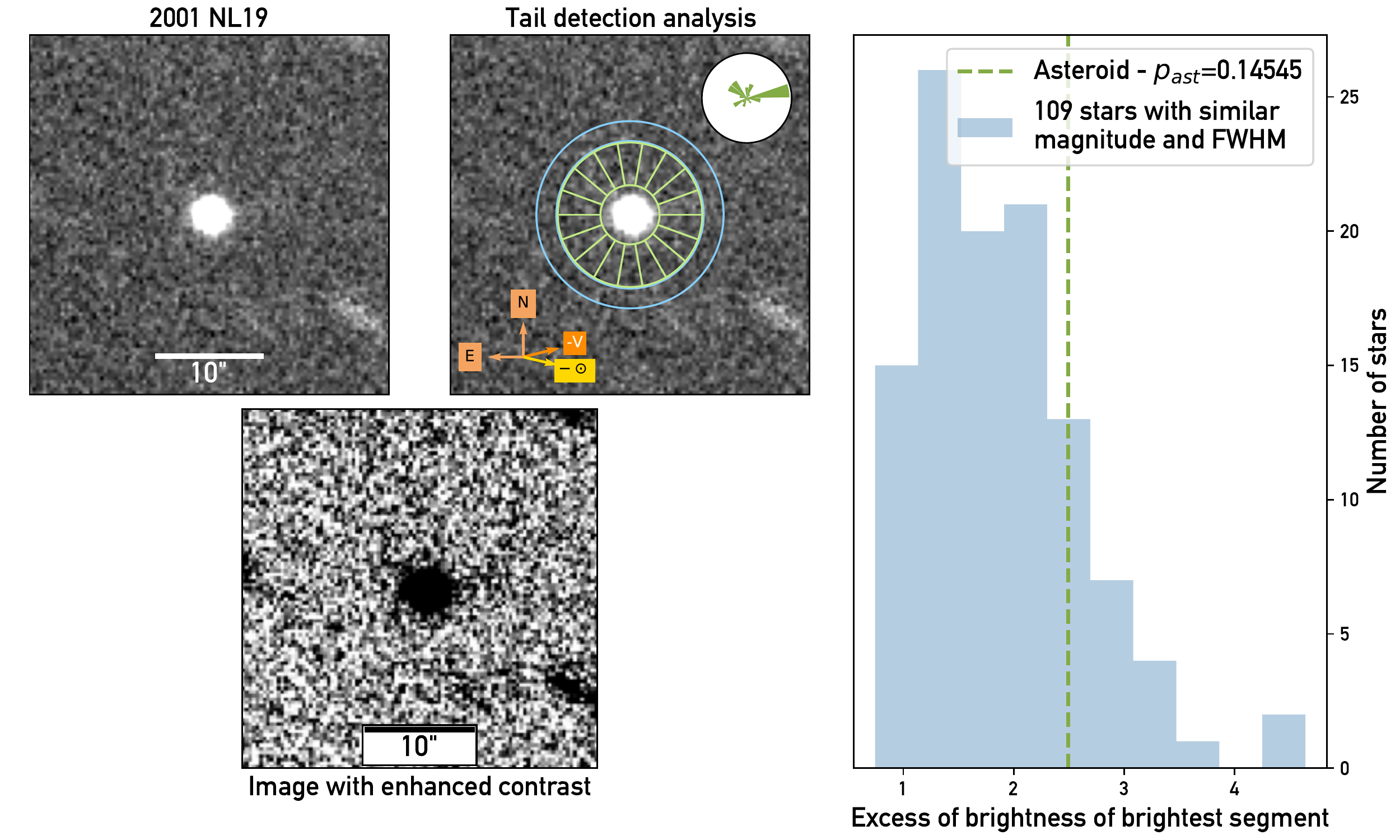}
    \caption{Result of the tail detection method for 2001 NL19 (279870). The object obtained a $p_{\mathrm{ast}}$ value of 0.18 and displays a faint tail-like feature in the west direction.}
    \label{fig:2001NL19}
\end{figure*}

2001 NL19 (279870) was observed on 2019 November 07, and obtained a $p_{\mathrm{ast}}$ value of 0.145 and a coma fraction of 0.014. Figure \ref{fig:2001NL19} shows the result of the tail analysis for 2001 NL19 as well as our deep image of the object with enhanced contrast.

279870 (2001 NL19) shows high levels of brightness in multiple segments in the west direction, which matches with the anti-velocity and anti-solar directions. A bright feature is visible in that region of the image. Although it does not unambiguously look like a tail, it could be a small faint tail or an outburst. 

PanSTARRS-1 images do not show a background source. Moreover, one should note that the direction of the feature is different from the trailing direction of the stars; the "tail" is horizontal while the stars appear trailed along a diagonal axis. If there was a background star behind the target it would likely produce a diagonal tail-like feature.
There is no visible contamination on any individual frame such as cosmic rays that could explain the bright feature on the stacked image.

2001 NL19 is also visible on multiple images from the Asteroid Terrestrial-impact Last Alert System (ATLAS) \citep{atlassurvey} between September 2019 and January 2020.  However the variations in magnitude and orientation among the images, as well as the expected decrease in activity during that period (2001 NL19 reached perihelion in May 2019) complicates the exploitation of the data to look for activity. We did not find any sign of activity in the ATLAS data, although our INT images are typically deeper than ATLAS ones.

We cannot definitely conclude on the activity of 2001 NL19 and further observations are needed to lift the ambiguity.

\section{Discussion}
\label{sec:discussion}

As explained in previous sections, our analysis suffers from multiple biases and limitations.  As we are dealing with a large amount of data with varying quality, density of stars, artefacts, etc., automating the data reduction and analysis was challenging. In some cases the alignment of frames or identification of the asteroid can lack accuracy. The detection of close neighbouring stars can be inaccurate, and some sets of frames had to be removed from or added to the results \textit{a posteriori}. This unreliability also introduces a bias in the results of tail detection and, to a lesser extent, can affect the calculation of asteroid models for coma detection. The modelling of the motion of the asteroids is sometimes imprecise which can influence coma detection as well.

We chose to review and correct the results manually rather than aiming for a fully-automated process.  
We would rather have to inspect the frames manually than have activity go unnoticed because it did not exactly fit stricter detection criteria.

Still, this analysis allowed us to identify one potentially active object among our sample. 2001 NL19 will be back at perihelion in October 2023 and we will conduct more observations of this object to look for activity. 

If 2001 NL19 is a Main Belt Comet, our study yields a detection rate of 1:316 if we do not account for the targets that could not be processed by the analysis pipeline and for which activity that is faint would most likely go undetected. In comparison, \citet{Sonnett2011} derived a MBC to asteroid ratio of $<1:500$ for the Outer Main Belt, meaning that our targeted survey would seem more efficient than non-targeted surveys. This would indicate the veracity and usefulness of the clustering of longitudes of perihelion found by \citet{kim}.
If this object is not active, then our study is consistent with previous limits put on the MBC population.

%______________________________
\section{Conclusion}
\label{sec:conclusion}

Using the WFC on the INT we imaged 534 asteroids with longitudes of perihelion close to that of Jupiter, in accordance with the hypothesis of \citet{kim}, making up a total of 554 sets of frames. We adapted activity detection methods developed by \citet{Sonnett2011} to look for tails or comae on our images, and were able to apply this analysis pipeline to 319 sets of frames, corresponding to 316 different asteroids.

Among the objects that were analysed we manually inspected the ones with the lowest $p_{\mathrm{ast}}$ values, which might indicate the presence of a tail. Most of these objects  did not seem active or reached low $p_{\mathrm{ast}}$ values due to contamination by faint background sources. However asteroid 2001 NL19 seems to show a faint bright feature matching the antivelocity and antisolar directions. This feature does not seem to be due to background sources or cosmic rays, and we are not aware of other deep observations of the object at that time to confirm activity. Further observations of the object near its next perihelion could help confirm its MBC nature, and the effectiveness of the target selection criterion.

This study also brought to light the many adjustments that had to be made to the original analysis method to fit the characteristics of WFC data products. A similar pipeline for data from other instruments is currently under development, in particular for data from the Vera C. Rubin Observatory.

%______________________________
\section*{Acknowledgements}

The authors would like to thank the anonymous referee for their valuable input on an earlier version of the manuscript.

This work began with discussions at the International Space Sciences Institute (ISSI), Bern, Switzerland, as part of the ISSI International Team on MBCs led by Colin Snodgrass. We thank ISSI for their hospitality and support.

This work has made use of data from the Asteroid Terrestrial-impact Last Alert System (ATLAS) project. ATLAS is primarily funded to search for near earth asteroids through NASA grants NN12AR55G, 80NSSC18K0284, and 80NSSC18K1575; byproducts of the NEO search include images and catalogues from the survey area. The ATLAS science products have been made possible through the contributions of the University of Hawaii Institute for Astronomy, the Queen's University Belfast, the Space Telescope Science Institute, and the South African Astronomical Observatory.

For the purpose of open access, the authors have applied a Creative Commons Attribution (CC BY) licence to any Author Accepted Manuscript version arising from this submission.

\section*{Data availability}

The data underlying this article will be shared on reasonable request to the corresponding author.

\bibliographystyle{mnras}
\bibliography{biblio} 

\begin{thebibliography}{}
\makeatletter
\relax
\def\mn@urlcharsother{\let\do\@makeother \do\$\do\&\do\#\do\^\do\_\do\%\do\~}
\def\mn@doi{\begingroup\mn@urlcharsother \@ifnextchar [ {\mn@doi@}
  {\mn@doi@[]}}
\def\mn@doi@[#1]#2{\def\@tempa{#1}\ifx\@tempa\@empty \href
  {http://dx.doi.org/#2} {doi:#2}\else \href {http://dx.doi.org/#2} {#1}\fi
  \endgroup}
\def\mn@eprint#1#2{\mn@eprint@#1:#2::\@nil}
\def\mn@eprint@arXiv#1{\href {http://arxiv.org/abs/#1} {{\tt arXiv:#1}}}
\def\mn@eprint@dblp#1{\href {http://dblp.uni-trier.de/rec/bibtex/#1.xml}
  {dblp:#1}}
\def\mn@eprint@#1:#2:#3:#4\@nil{\def\@tempa {#1}\def\@tempb {#2}\def\@tempc
  {#3}\ifx \@tempc \@empty \let \@tempc \@tempb \let \@tempb \@tempa \fi \ifx
  \@tempb \@empty \def\@tempb {arXiv}\fi \@ifundefined
  {mn@eprint@\@tempb}{\@tempb:\@tempc}{\expandafter \expandafter \csname
  mn@eprint@\@tempb\endcsname \expandafter{\@tempc}}}

\bibitem[\protect\citeauthoryear{Barbary}{Barbary}{2016}]{Barbary2016}
Barbary K.,  2016, \mn@doi [Journal of Open Source Software]
  {10.21105/joss.00058}, 1, 58

\bibitem[\protect\citeauthoryear{Beroiz, Cabral  \& Sanchez}{Beroiz
  et~al.}{2020}]{Beroiz2020}
Beroiz M.,  Cabral J.,   Sanchez B.,  2020, \mn@doi [Astronomy and Computing]
  {https://doi.org/10.1016/j.ascom.2020.100384}, 32, 100384

\bibitem[\protect\citeauthoryear{{Berthier}, {Vachier}, {Thuillot}, {Fernique},
  {Ochsenbein}, {Genova}, {Lainey}  \& {Arlot}}{{Berthier}
  et~al.}{2006}]{skybot}
{Berthier} J.,  {Vachier} F.,  {Thuillot} W.,  {Fernique} P.,  {Ochsenbein} F.,
   {Genova} F.,  {Lainey} V.,   {Arlot} J.~E.,  2006, {SkyBoT, a new VO service
  to identify Solar System objects}

\bibitem[\protect\citeauthoryear{{Bertin} \& {Arnouts}}{{Bertin} \&
  {Arnouts}}{1996}]{sextractor}
{Bertin} E.,  {Arnouts} S.,  1996, \mn@doi [\aaps] {10.1051/aas:1996164}, \href
  {https://ui.adsabs.harvard.edu/abs/1996A&AS..117..393B} {117, 393}

\bibitem[\protect\citeauthoryear{Elst, Pizarro, Pollas, Ticha, Tichy, Moravec,
  Offutt  \& Marsden}{Elst et~al.}{1996}]{Elst}
Elst E.~W.,  Pizarro O.,  Pollas C.,  Ticha J.,  Tichy M.,  Moravec Z.,  Offutt
  W.,   Marsden B.~G.,  1996, IAU Circ. 6456

\bibitem[\protect\citeauthoryear{{Exolab Group}}{{Exolab
  Group}}{2002}]{openorb}
{Exolab Group} 2002, The OpenORB project.
\url {http://openorb.exolab.org/}

\bibitem[\protect\citeauthoryear{Gilbert \& Wiegert}{Gilbert \&
  Wiegert}{2009}]{GILBERT2009}
Gilbert A.~M.,  Wiegert P.~A.,  2009, \mn@doi [Icarus]
  {https://doi.org/10.1016/j.icarus.2009.01.011}, 201, 714

\bibitem[\protect\citeauthoryear{Gilbert \& Wiegert}{Gilbert \&
  Wiegert}{2010}]{GILBERT2010}
Gilbert A.~M.,  Wiegert P.~A.,  2010, \mn@doi [Icarus]
  {https://doi.org/10.1016/j.icarus.2010.07.016}, 210, 998

\bibitem[\protect\citeauthoryear{Hsieh, Schwamb, Zhang, Chen, Wang  \&
  Lintott}{Hsieh et~al.}{2016}]{Hsieh2016}
Hsieh H.,  Schwamb M.,  Zhang Z.,  Chen Y.,  Wang S.,   Lintott C.,  2016,
  American Astronomical Society/Division for planetary sciences meeting
  abstracts, 48, 406 03

\bibitem[\protect\citeauthoryear{{Jewitt} \& {Hsieh}}{{Jewitt} \&
  {Hsieh}}{2022}]{JewittHsieh2022}
{Jewitt} D.,  {Hsieh} H.~H.,  2022, arXiv e-prints, \href
  {https://ui.adsabs.harvard.edu/abs/2022arXiv220301397J} {p. arXiv:2203.01397}

\bibitem[\protect\citeauthoryear{{Kim}, {JeongAhn}  \& {Hsieh}}{{Kim}
  et~al.}{2018}]{kim}
{Kim} Y.,  {JeongAhn} Y.,   {Hsieh} H.~H.,  2018, \mn@doi [\aj]
  {10.3847/1538-3881/aaad01}, \href
  {https://ui.adsabs.harvard.edu/abs/2018AJ....155..142K} {155, 142}

\bibitem[\protect\citeauthoryear{{Lang}, {Hogg}, {Mierle}, {Blanton}  \&
  {Roweis}}{{Lang} et~al.}{2010}]{Lang2010}
{Lang} D.,  {Hogg} D.~W.,  {Mierle} K.,  {Blanton} M.,   {Roweis} S.,  2010,
  \mn@doi [\aj] {10.1088/0004-6256/139/5/1782}, \href
  {https://ui.adsabs.harvard.edu/abs/2010AJ....139.1782L} {139, 1782}

\bibitem[\protect\citeauthoryear{Schwamb, Hsieh, Zhang, Chen, Lintott, Wang  \&
  Mishra}{Schwamb et~al.}{2017}]{Schwamb2017}
Schwamb M.,  Hsieh H.,  Zhang Z.,  Chen Y.,  Lintott C.,  Wang S.,   Mishra I.,
   2017, American Astronomical Society meeting abstracts, 229, 112 04

\bibitem[\protect\citeauthoryear{{Snodgrass}, {Lowry}  \&
  {Fitzsimmons}}{{Snodgrass} et~al.}{2008}]{snodgrass2007}
{Snodgrass} C.,  {Lowry} S.~C.,   {Fitzsimmons} A.,  2008, \mn@doi [\mnras]
  {10.1111/j.1365-2966.2008.12900.x}, \href
  {https://ui.adsabs.harvard.edu/abs/2008MNRAS.385..737S} {385, 737}

\bibitem[\protect\citeauthoryear{Snodgrass et~al.,}{Snodgrass
  et~al.}{2017}]{Snodgrass2017}
Snodgrass C.,  et~al., 2017, \mn@doi [Astronomy and Astrophysics Review]
  {10.1007/s00159-017-0104-7}, 25

\bibitem[\protect\citeauthoryear{Sonnett, Kleyna, Jedicke  \& Masiero}{Sonnett
  et~al.}{2011}]{Sonnett2011}
Sonnett S.,  Kleyna J.,  Jedicke R.,   Masiero J.,  2011, \mn@doi [Icarus]
  {10.1016/j.icarus.2011.08.001}, 215, 534

\bibitem[\protect\citeauthoryear{{Tonry} et~al.,}{{Tonry}
  et~al.}{2018}]{atlassurvey}
{Tonry} J.~L.,  et~al., 2018, \mn@doi [\pasp] {10.1088/1538-3873/aabadf}, \href
  {https://ui.adsabs.harvard.edu/abs/2018PASP..130f4505T} {130, 064505}

\bibitem[\protect\citeauthoryear{Waszczak et~al.,}{Waszczak
  et~al.}{2013}]{Waszczak}
Waszczak A.,  et~al., 2013, \mn@doi [\mnras] {10.1093/mnras/stt951}, 433, 3115

\makeatother
\end{thebibliography}

% if your bibtex file is called example.bib

%\begin{thebibliography}{99}
%\end{thebibliography}

%%%%%%%%%%%%%%%%%%%%%%%%%%%%%%%%%%%%%%%%%%%%%%%%%%

%%%%%%%%%%%%%%%%% APPENDICES %%%%%%%%%%%%%%%%%%%%%
\onecolumn
\newpage
\appendix

\section{Full observation log}
\label{appendixlog}

\begin{table}
\centering
\footnotesize
\caption{Extract of the full observation log, which includes observing parameters, orbital characteristics of each target and results of the tail and coma detection methods. $\Delta$: Observer distance; R$_{h}$: Heliocentric distance; STO: Sun-Target-Observer angle; a: Semi-major axis; e: Eccentricity; i: Inclination; T$_{J}$: Tisserand parameter with respect to Jupiter;  $\Delta t_{\mathrm{peri}}$: Time to perihelion; V: Apparent magnitude; H: Absolute magnitude; $\nu$: True anomaly; $p_{\mathrm{ast}}$: Tail detection indicator; $f_{\mathrm{c}}$: Coma fraction. The full table will be made available online.}
\begin{tabular}{rcc|c|ccccccc}
\hline
UT time & & Exposures (s) & Object & $\Delta$ (AU) & R$_{h}$ (AU) & STO (deg) & a (AU) & e & i (deg) & T$_{J}$ \\ \hline 
2018-11-07 &  19:30:40.5 & 10$\times$ 50 & 426856 & 1.98 & 2.426 & 23.3 & 3.13 & 0.229 & 16.4 & 3.11 \\ %1
2018-11-07 &  19:45:52.5 & 10$\times$ 50 & 378471 & 1.991 & 2.399 & 23.8 & 3.15 & 0.241 & 9.32 & 3.14  \\ %2
2018-11-07 &  19:59:56.5 & 10$\times$ 50 & 2002 TU167 & 1.777 & 2.262 & 24.9 & 3.07 & 0.269 & 18.6 &  3.1 \\ %3
2018-11-07 &  20:14:08.9 & 7$\times$ 71 & 464725 & 1.707 & 2.284 & 23.5 & 3.07 & 0.257 & 8.64 & 3.16 \\ %4
2018-11-07 &  20:26:42.5 & 6$\times$ 83 & 2001 RB105 & 1.766 & 2.351 & 22.6 & 3.16 & 0.257 & 9.19 &   1.70\\ %5
2018-11-07 &  20:38:51.5 & 7$\times$ 71 & 276362 & 1.86 & 2.466 & 21.1 & 3.1 & 0.205 & 2.02 & 3.19  \\ %6
2018-11-07 &  20:51:27.5 & 8$\times$ 62 & 2007 SH20 & 1.896 & 2.427 & 22.5 & 3.15 & 0.23 & 10.7 & 3.14   \\ %7
2018-11-07 &  21:04:44.5 & 7$\times$ 71 & 332435 & 1.773 & 2.422 & 20.8 & 3.18 & 0.241 & 14.0 & 3.11 \\ %8
2018-11-07 &  21:17:07.5 & 3$\times$ 166 & 193897 & 1.656 & 2.391 & 19.2 & 3.15 & 0.243 & 8.38 & 3.15  \\ %9
2018-11-07 &  21:27:50.5 & 3$\times$ 166 & 113556 & 1.932 & 2.574 & 19.5 & 3.08 & 0.167 & 1.68 & 3.21 \\ %10
 &  &  &  & [...] & & & &  &  \\ \hline %10
UT time & & Exposures (s) & Object & $\bar{\omega}$ (deg) & $\Delta t_{peri}$ (d) & V & H & $\nu$ (deg) & $p_{\mathrm{ast}}$ & $f_{\mathrm{c}}$ \\ \hline 
 2018-11-07 &  19:30:40.5 & 10$\times$50 & 426856 & 2.0 & 28.8 & 20.6 & 16.1 & 351.1 & 0.5 & 1.2E-4 \\ %1
2018-11-07 &  19:45:52.5 & 10$\times$50 & 378471 & 2.2 & 36.5 & 20.8 & 16.3 & 349.4 & N/A & N/A \\ %2
2018-11-07 &  19:59:56.5 & 10$\times$50 & 2002 TU167 & 11.0 & 47.4 & 20.2 & 16.1 & 344.9 & N/A & N/A \\ %3
2018-11-07 &  20:14:08.9 & 7$\times$71 & 464725 & 3.5 & 7.8 & 20.4 & 16.3 & 358.8 & 0.84 & 1.0E-4 \\ %4
2018-11-07 &  20:26:42.5 & 6$\times$83 & 2001 RB105 & 4.5 & 9.3 & 21.0 & 16.8 & 357.9 & 0.1 & 2.0E-5 \\ %5
2018-11-07 &  20:38:51.5 & 7$\times$71 & 276362 & 1.3 & -3.4 & 20.5 & 16.1 & 1.3 & N/A & N/A \\ %6
2018-11-07 &  20:51:27.5 & 8$\times$62 & 2007 SH20 & 1.0 & 11.1 & 21.1 & 16.7 & 356.9 & 0.31 & 8.4E-11 \\ %7
2018-11-07 &  21:04:44.5 & 7$\times$71 & 332435 & 10.0 & 12.0 & 20.0 & 15.8 & 355.4 & 0.9 & 1.1E-4 \\ %8
2018-11-07 &  21:17:07.5 & 3$\times$166 & 193897 & 12.8 & 5.8 & 19.3 & 15.3 & 359.5 & N/A & N/A \\ %9
2018-11-07 &  21:27:50.5 & 3$\times$166 & 113556 & 5.0 & -0.1 & 19.8 & 15.3 & 360.0 & N/A & N/A \\ %10
 &  &  &  & [...] & & & &  &  \\ \hline %10

\end{tabular}

\label{tab:obslog}
\end{table}

% Don't change these lines
\bsp	% typesetting comment
\label{lastpage}
\end{document}